\documentclass[aps, prd, groupedaddress, preprint, eqsecnum, nofootinbib, showpacs]{revtex4}
\usepackage{graphicx,epsf,amssymb,amsbsy,amsfonts,amssymb,amsmath,mathrsfs}
\usepackage[usenames, dvipsnames]{color}
\usepackage{pstricks}
\def\ka{\kappa}
\def\tree{{\rm tree}}
\def\oneloop{{(1)}} 
\def\twoloop{{(2)}} 
\def\Lloop{{(L)}}

\def\e{\epsilon}
\def\Tr{\, {\rm Tr}}

\def\nn{\nonumber}

\def\I{{\cal I}}

\def\f{\tilde f}
\def\polylog{ {\rm Li} }
\def\lr{\leftrightarrow}

\def\NeqSix{{{\cal N}=6}}
\def\NeqFive{{{\cal N}=5}}
\def\NeqFour{{{\cal N}=4}}
\def\NeqTwo{{{\cal N}=2}}
\def\NeqZero{{{\cal N}=0}}

\def\NeqEight{{{\cal N}=8}}

\def\NeqOne{{{\cal N}=1}}
\def\NYM{{\cal N}_{\rm YM}}

\def\Ord{{\cal O}}
\def\P{({\rm P})}
\def\NP{({\rm NP})}

\def\spa#1.#2{\left\langle#1\,#2\right\rangle}
\def\spb#1.#2{\left[#1\,#2\right]}
\DeclareMathAlphabet{\mathpzc}{OT1}{pzc}{m}{it}

\newif\ifdraft
\drafttrue
\newif\ifpreprint
\preprinttrue

\def\be{\begin{equation}}
\def\ee{\end{equation}}
\def\bea{\begin{eqnarray}}
\def\eea{\end{eqnarray}}
\def\ba{\begin{eqnarray}}
\def\ea{\end{eqnarray}}
\def\sect#1{section~{\ref{#1}}}
\def\fig#1{fig.~{\ref{#1}}}

\def\app#1{appendix~{\ref{#1}}}
\def\eqn#1{eq.~(\ref{#1})}

\def\eqns#1#2{eqs.~(\ref{#1}) and~(\ref{#2})}

\def\tab#1{table~{\ref{#1}}}

\begin{document}
\hfuzz 20pt

\ifpreprint
\begin{flushright}
SLAC-PUB-14550
\end{flushright}                        
\fi

\title{${\cal N} \ge 4$ Supergravity Amplitudes from Gauge Theory
at Two Loops}

\author{C.~Boucher-Veronneau and L.~J.~Dixon}

\vskip0.5cm

\affiliation{
SLAC National Accelerator Laboratory,
Stanford University, Stanford, CA 94309, USA \\
$\null$\\
$\null$\\
}

\begin{abstract}
We present the full two-loop four-graviton amplitudes in
${\cal N}=4,5,6$ supergravity. These results were obtained using the
double-copy structure of gravity, which follows from the recently
conjectured color-kinematics duality in gauge theory.
The two-loop four-gluon scattering
amplitudes in ${\cal N}=0,1,2$ supersymmetric gauge theory are a
second essential ingredient.  The gravity amplitudes have the expected
infrared behavior: the two-loop divergences are given in terms of the
squares of the corresponding one-loop amplitudes.  The finite remainders
are presented in a compact form.  The finite remainder for $\NeqEight$
supergravity is also presented, in a form that utilizes a pure function
with a very simple symbol.
\end{abstract}

\pacs{04.65.+e, 11.15.Bt, 11.30.Pb, 11.55.Bq \hspace{1cm}}

\maketitle


\section{Introduction}

It is well known that pure Einstein gravity is ultraviolet (UV) divergent
at two loops~\cite{Einstein2Loop}. This result, along with general
power-counting arguments, has led to the widespread belief that a UV finite
pointlike theory of gravity cannot be constructed. However, explicit
calculations of scattering amplitudes in maximally supersymmetric
($\NeqEight$) supergravity have displayed an ultraviolet behavior that is
much better than prior expectations, showing that the theory in four
dimensions is finite up to at least four loops.  Furthermore, $\NeqEight$
supergravity exhibits the same UV behavior, when continued to higher
spacetime dimensions, as does $\NeqFour$ super-Yang-Mills
(sYM)~\cite{FourLoopNeq8,FourLoop,FourLoopBCJ}.  Surprising cancellations
are also visible at lower loop
orders~\cite{TwoLoop,ThreeLoop,Multileg,Cancellations,AHCK,FivePointBCJ},
and even at tree level where the amplitudes are nicely behaved at large
(complex) momenta~\cite{TreeCancel,AHCK}.

In pure supergravity theories (where all states are related by
supersymmetry to the graviton) no counterterm can be constructed below
three loops. This is because the only possible two-loop counterterm,
$R^3 \equiv R^{\lambda\rho}_{\mu\nu} R^{\mu\nu}_{\sigma\tau}
R^{\sigma\tau}_{\lambda\rho}$,
where $R^{\mu\nu}_{\sigma\tau}$ is the
Riemann tensor, generates non-zero four-graviton amplitudes with helicity
assignment $(\pm,+,+,+)$~\cite{Grisaru,DKS,Tomboulis}.  Such amplitudes
are forbidden by the Ward identities for the minimal $\NeqOne$
supersymmetry~\cite{SUSYWard}.  The counterterm denoted by $R^4$ is
allowed by supersymmetry and could appear at three loops~\cite{DKS,R4ct}.
However, as mentioned earlier, $\NeqEight$ supergravity was found to be
finite at this order~\cite{ThreeLoop}.  It was recently understood that
the $R^4$ counterterm is forbidden~\cite{BDR4,EKR4} by the nonlinear
$E_{7(7)}$ symmetry realized by the 70 scalars of the
theory~\cite{CremmerJulia,Bossard2010dq}.  In fact, $E_{7(7)}$ should
delay the divergence in $\NeqEight$ supergravity to at least seven loops,
where the first $E_{7(7)}$-invariant counterterm can be
constructed~\cite{BeisertetalR4,BHS,BHSV}.  Non-maximal (${\cal N} < 8$)
supergravity does not have this extra $E_{7(7)}$ symmetry, and may
therefore diverge at only three loops in four dimensions.

Recently, the constraints that the smaller duality symmetries of non-maximal
supergravities impose on potential counterterms have also been
investigated~\cite{BHS,BHSV}.  In four dimensions,
$\NeqSix$ supergravity is expected to be finite at three and four loops,
and $\NeqFive$ supergravity should be finite at three loops~\cite{BHS}.
These results still allow for a three-loop divergence in ${\cal N}\leq 4$
supergravities.  In particular, for $\NeqFour$ supergravity,
although the volume of superspace vanishes on shell, it has been argued
that the usual three-loop $R^4$ counterterm can appear~\cite{BHSV}.
The finiteness results for ${\cal N}=5,6$ could in principle be checked,
and potential divergences for ${\cal N}\leq 4$ investigated, via explicit
three-loop amplitude calculations in non-maximal supergravities.
Because the same situation, in which the superspace volume vanishes
on shell, and yet a counterterm appears to be allowed, holds for
$\NeqEight$ supergravity at seven loops, as for $\NeqFour$ supergravity
at three loops, this latter case may be of particular interest.

On the other hand, relatively few loop amplitudes have been computed 
for any non-maximal supergravities. At one loop, the four-point
amplitudes with ${\cal N} \le 8$ supersymmetries were presented in
ref.~\cite{DunbarNorridge}, while the $\NeqSix$ supergravity all-point
maximally-helicity-violating (MHV)
and six-point non-MHV amplitudes were first obtained in
ref.~\cite{DunbarFivePts}. The $\NeqFour$ supergravity
one-loop five-point amplitude was also computed in
refs.~\cite{DunbarFivePts,Nle8BCJ}.
In the following, we present expressions for the two-loop four-graviton
amplitudes in ${\cal N}=4,5,6$ supergravity. The calculations were
performed using the gravity ``squaring"
relations~\cite{BCJLoop,Bern2010yg}, or double-copy property,
which follows from the color-kinematics, or Bern-Carrasco-Johansson
(BCJ), duality obeyed by gauge-theory amplitudes at the loop
level~\cite{Bern2008qj}.

The BCJ relations allow us to combine the $\NeqFour$ sYM
amplitude~\cite{TwoLoopYMn4} with the ${\cal N} = 0,1,2$ sYM
amplitudes~\cite{TwoLoopYMn2} in order to obtain the
corresponding amplitudes in supergravity.
Although they have been tested now in several loop-level
amplitude computations~\cite{Bern2008qj,BCJLoop,FivePointBCJ,FourLoopBCJ},
the underlying mechanism or symmetry behind the general loop-level BCJ
relations is still not well understood.  (In the self-dual sector
at tree level, a diffeomorphism Lie algebra appears to play
a key role.~\cite{MOC}.)  Therefore it is important to
validate results obtained using BCJ duality.  We will verify
the expected infrared divergences and forward-scattering behavior
for the two-loop amplitudes that we compute.

This paper is organized as follows. In~\sect{BCJsquarereview} we 
review BCJ duality and the squaring relations for gravity.
In~\sect{Neq8warmup} we illustrate the method for $\NeqEight$ supergravity
at two loops.  In~\sect{Neq456masterformula} we present our main
formula for the two-loop amplitudes in ${\cal N} =4,5,6$ supergravity.
In~\sect{IRandFiniteSection} we expand the (dimensionally regulated)
amplitudes for $D=4-2\e$ around $\e=0$.  We discuss the infrared (IR) pole
structure, which agrees with general expectations, thus providing
a cross check on the construction.  We present the
finite remainders in the two independent kinematic channels.
In~\sect{ForwardSection} we examine the behavior of the amplitudes 
in the limit of forward scattering.
In~\sect{ConclusionSection}, we present our conclusions
and suggestions for future research directions.
An appendix provides some one-loop results that are required for extracting
the two-loop finite remainders.


\section{Review of the BCJ duality and squaring relations}
\label{BCJsquarereview}

We now briefly review BCJ duality and the gravity squaring relations
that follow from it. For a more complete treatment see, for example, the
recent reviews~\cite{BCJReviewCJ,BCJReviewS}. Here, we will focus solely on
applications to loop amplitudes.

We can write any $m$-point $L$-loop-level gauge-theory amplitude, where all
particles are in the adjoint representation, as
\begin{equation}
 {\cal A}^{\Lloop}_m\ =\ i^L \, g^{m - 2 + 2 L} \,
 \sum_{j}{\int \prod_{l=1}^L \frac{ d^{D} p_l}{(2 \pi)^{D}}
  \frac{1}{S_j}  \frac {n_j c_j}{\prod_{\alpha_j}{p^2_{\alpha_j}}}}\,,
\label{LoopGauge} 
\end{equation}
where $g$ is the gauge coupling.  The sum runs over
the set of distinct $m$-point $L$-loop graphs, labeled by $j$, with
only cubic vertices, corresponding to the diagrams of a $\phi^3$
theory.  The product in the denominator runs over all Feynman
propagators of each cubic diagram.  The integrals are over $\{p_l^\mu\}$,
a set of $L$ independent $D$-dimensional loop momenta.
The $c_i$ are the color factors, obtained by
dressing every three-vertex with a structure constant, defined by
$\f^{abc} = i \sqrt{2} f^{abc}=\Tr\bigl( [T^{a}, T^{b}] T^{c} \bigr)$.
The $n_j$ are kinematic numerator factors depending on momenta, polarizations
and spinors.  The $S_j$ are the internal symmetry factors for each diagram.
The form of the amplitude presented in~\eqn{LoopGauge} can be obtained in
various ways.  For example, one can start from covariant Feynman diagrams
in Feynman gauge, where the contact terms are absorbed into kinematic
numerators using inverse propagators, {\it i.e.}~by inserting factors of
$1 = p_{\alpha_j}^2/p_{\alpha_j}^2$.

Triplets $(i,j,k)$ of color factors are related to each other by
$c_i = c_j + c_k$ if their corresponding graphs are identical,
except for a region containing (in turn for $i,j,k$) the three cubic
four-point graphs that exist at tree level.  The relation holds because
the products of two $\f^{abc}$ structure constants
corresponding to the four-point tree graphs satisfy the Jacobi identity
\be
\f^{abe} \, \f^{cde}\ =\ \f^{ace} \, \f^{bde} + \f^{ade} \, \f^{cbe} \,,
\label{ffJacobi}
\ee
and the remaining structure constant factors in the triplet of graphs
are identical.  The relations $c_i = c_j + c_k$ mean that the
representation~(\ref{LoopGauge}) is not unique; terms can be shuffled from
one graph to others, in a kind of generalized gauge
transformation~\cite{BCJLoop}.

A representation~(\ref{LoopGauge}) is said to satisfy the BCJ duality if the
three associated kinematic numerators are also related via Jacobi
identities. Namely, we must have:
 \begin{equation} \label{jacobi}
 c_i = c_j + c_k  \qquad  \Rightarrow \qquad  n_i = n_j + n_k \,,
 \end{equation}
where the left-hand side follows directly from group theory, while the
right-hand side is the highly non-trivial requirement of the
duality. Moreover, we demand that the numerator factors have the same
antisymmetry property as the color factors under the interchange of two legs
attached to a cubic vertex,
\begin{equation}
c_i \rightarrow - c_i \qquad  \Rightarrow \qquad n_i \rightarrow - n_i \,.
\label{BCJAntiSymmetry}
\end{equation}
The relations~(\ref{jacobi}) were found long ago for the case of four-point
tree amplitudes~\cite{treeBCJ4pt}; the idea that the relations should
hold for arbitrary amplitudes is more recent~\cite{Bern2008qj,BCJLoop}.

As remarked earlier, the representation~(\ref{LoopGauge}) is not
unique.  Work is often required in order to find a BCJ-satisfying
representation of a given amplitude in a particular gauge theory.
At loop level, such representations were found initially at four points
through three loops for $\NeqFour$ sYM,
and through two loops for identical-helicity pure
Yang-Mills amplitudes~\cite{BCJLoop}.  A BCJ-satisfying
representation was recently obtained at five points through three loops in
$\NeqFour$ sYM~\cite{FivePointBCJ}. Very recently, a four-point four-loop
representation was found in the same theory~\cite{FourLoopBCJ}.

As a remarkable consequence of the BCJ duality, one can combine two
gauge-theory amplitudes in the form~(\ref{LoopGauge}), in order to obtain a
gravity amplitude, as long as one of the two gauge-theory representations
manifestly satisfies the duality~\cite{BCJLoop,Bern2010yg}. 
We have,
\begin{equation}\label{bcj_gravity}
{\cal M}_m^{\Lloop}
\ =\ i^{L+1} \, \left( \frac{\ka}{2} \right)^{m - 2 + 2 L} \,
 \sum_j \int \prod_{l=1}^{L} \frac{d^Dp_l}{(2\pi)^D}
\frac{1}{S_j}\frac{n_j\tilde{n}_j}{\prod_{\alpha_j} p_{\alpha_j}^2} \,,
\end{equation}
where either the $n_j$ or the $\tilde{n}_j$ must satisfy
\eqns{jacobi}{BCJAntiSymmetry}.
Here $\ka$ is the gravitational coupling constant, which is
related to Newton's constant $G_N$ and the Planck mass
$M_{\rm Planck}$ by $\ka^2 = 32\pi G_N = 32\pi/M_{\rm Planck}^2$.
The proof of \eqn{bcj_gravity} at tree level is inductive, and uses on-shell
recursion relations~\cite{BCFW} for the gauge and gravity theories,
which are based on the same complex momentum shift~\cite{Bern2010yg}.
The extrapolation to loop level is based on reconstructing loop amplitudes
from tree amplitudes using (generalized) unitarity.

The relations~(\ref{bcj_gravity}) are similar in spirit
to the KLT relations~\cite{KLT}.  Both types of relations express gravity
amplitudes as the ``square'' of gauge-theory amplitudes, or more generally
as the product of two different types of gauge-theory amplitudes, as
the $n_i$ and $\tilde{n}_j$ numerator factors may come from two different
Yang-Mills theories.  However, the KLT relations only hold
at tree level, which means that at loop level they can only be used
on the (generalized) unitarity cuts.  Although the gravity cuts can be
completely determined by the KLT relations in terms of local Yang-Mills
integrands, the gravity integrand found in this way is not manifestly
local.  That is, it does not manifestly have the form of numerator factors
multiplied by scalar propagators for some set of $\phi^3$ graphs.
Reconstructing a local representation can be a significant
task~\cite{ThreeLoop,FourLoopNeq8}.

In contrast, \eqn{bcj_gravity} is a loop-level relation, and furnishes
directly a local integrand for gravity.  Most of the applications of
this formula to date have been to maximal $\NeqEight$ supergravity,
viewed as the tensor product of two copies of maximal $\NeqFour$
super-Yang-Mills theory.
The squaring relations were shown to reproduce the $\NeqEight$
supergravity four-point amplitudes through four
loops~\cite{BCJLoop,FourLoopBCJ} and the five-point amplitudes through two
loops~\cite{FivePointBCJ}.  Quite recently, in the first loop-level
applications for ${\cal N} < 8$, the one-loop four- and five-point
${\cal N} \le 8$ supergravity amplitudes were shown to satisfy the
double-copy property~\cite{Nle8BCJ}.  In this paper, we would like to
extend this kind of analysis for ${\cal N} < 8$ supergravity to two loops.
First, however, we briefly review the $\NeqEight$ case.


\section{Two-loop $\NeqEight$ supergravity}
\label{Neq8warmup}


\begin{figure*}[tb] 
\begin{center}
\includegraphics[scale=0.6]{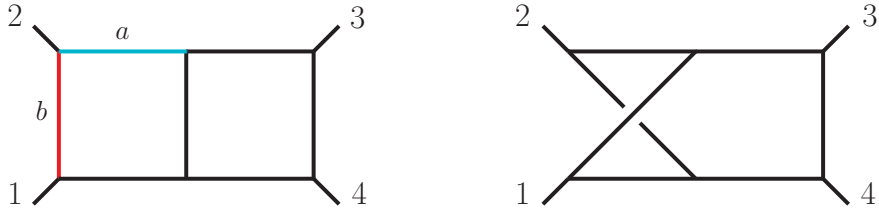}
\end{center}
\vskip -.9 cm
\caption[a]{\small The planar and nonplanar cubic diagrams at two loops.
The marked (colored) propagators in the planar diagram are used in the
text to describe different color and kinematic Jacobi identities.}
\label{TwoloopIntegralsFigure}
\end{figure*}


In this section we review the construction of the two-loop
four-graviton amplitude in $\NeqEight$ supergravity based on
squaring relations, as preparation for a similar construction
for ${\cal N}=4,5,6$ supergravity in the next section.

As mentioned previously, a manifestly BCJ-satisfying representation
of the four-gluon $\NeqFour$ sYM amplitude is known at two
loops~\cite{TwoLoopYMn4,BCJLoop}, 
\begin{eqnarray}\label{2loopNeq4ym}
{\cal A}^{\twoloop}_4(1,2,3,4) &=&  - g^6 st  \,
              A_4^{\tree}(1,2,3,4) \Bigl(
    c^{\P}_{1234} \, s \, \I_4^{\P}(s, t)
  + c^{\P}_{3421} \, s \, \I_4^{\P}(s, u) \\
 && \null \hskip  3.2 truecm
  + c^{\NP}_{1234} \, s\, \I_4^{\NP}(s,t)
  + c^{\NP}_{3421} \, s\, \I_4^{\NP}(s,u)
+  {\rm cyclic} \Bigr)\,, \nn
\end{eqnarray}
where $s,t,u$ are the usual Mandelstam invariants 
($s = (k_1 + k_2)^2$, $t = (k_2 + k_3)^2$, $u = (k_1 + k_3)^2$) 
and ``+ cyclic'' instructs one to add the two cyclic permutations of
(2,3,4).  The tree-level partial amplitude is
\be
A_4^\tree(1,2,3,4) = i \,
\frac{{\spa{j}.{k}}^4}{\spa1.2\spa2.3\spa3.4\spa4.1} \,,
\label{tree4pt}
\ee
where $j$ and $k$ label the two negative-helicity gluons.
The two-loop planar and nonplanar scalar double-box
integrals are, respectively,
\begin{eqnarray}
\I_4^{\P}(s,t) &=& \int
 \frac{d^{D}p}{ (2\pi)^{D}} 
 \frac{d^{D}q}{ (2\pi)^{D}} 
\frac {1}{ p^2 \, (p - k_1)^2 \,(p - k_1 - k_2)^2 \,(p + q)^2 q^2 \,
        (q-k_4)^2 \, (q - k_3 - k_4)^2 } \,, \nn \\
\I_4^{\NP}(s,t) &=& \int \frac{d^{D} p} { (2\pi)^{D}} \, 
           \frac {d^{D} q }{ (2\pi)^{D}} \
\frac{1}{ p^2\, (p-k_2)^2 \,(p+q)^2 \,(p+q+k_1)^2\,
  q^2 \, (q-k_3)^2 \, (q-k_3-k_4)^2}\, ,\nn
\end{eqnarray}
and they are depicted in~\fig{TwoloopIntegralsFigure}. The color factors
$c^{({\rm P,NP})}_{ijkl}$ are obtained by dressing each vertex of
the associated diagram with a factor of $\f^{abc}$, and each internal
line with a $\delta_{ab}$.  All helicity information is
encoded in the prefactor $s t A^{\tree}_4(1,2,3,4)$, which is invariant
under all permutations, thanks to a Ward identity for $\NeqFour$
supersymmetry.

%
\begin{figure}[tb]
\begin{center}
\includegraphics[scale=0.55]{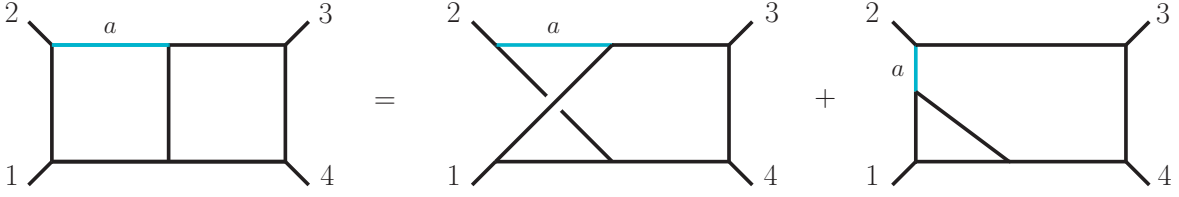}
\end{center}
\vskip -.9 cm 
\caption[a]{\small Two-loop diagrams related by a Jacobi identity. 
The Jacobi identity is applied to the four-point tree-level subdiagram
that contains the (light blue) intermediate line marked $a$.
The rest of the diagram is unchanged.}
\label{JacobiTwoLoopFigure}
\end{figure}
%

Comparing \eqns{LoopGauge}{2loopNeq4ym} we can identify the numerators as
\begin{eqnarray} 
 n^{\P}_{1234} = n^{\P}_{3421} = n^{\NP}_{1234} =
 n^{\NP}_{3421}&=& s \times s t \, A^{\tree}_4(1,2,3,4)\,, \nn \\
 n^{\P}_{1342} = n^{\P}_{4231} = n^{\NP}_{1342} =
 n^{\NP}_{4231}&=& u \times s t \, A^{\tree}_4(1,2,3,4)\,, \nn \\
 n^{\P}_{1423} = n^{\P}_{2341} = n^{\NP}_{1423} =
 n^{\NP}_{2341}&=& t \times s t \, A^{\tree}_4(1,2,3,4)\,.
\label{Neq4ymNums}
\end{eqnarray}
It is easy to see that the two-loop expression~(\ref{2loopNeq4ym})
satisfies the duality~\cite{BCJLoop}. For instance, let's look at the
diagrams related by a Jacobi identity applied to a four-point
tree-level subdiagram of the planar double-box graph on the
left-hand side of~\fig{TwoloopIntegralsFigure}.  The tree subdiagram
is the one whose intermediate propagator is the light-blue line marked
$a$ in the figure.  We replace the ``$s$-channel'' tree subdiagram with
the corresponding $t$- and $u$-channel tree subdiagrams, by appropriately
permuting the attachments of line $a$ to the rest of the graph.
This Jacobi identity is illustrated in~\fig{JacobiTwoLoopFigure}.
Because the $\NeqFour$ sYM diagrams with triangle one-loop subdiagrams
all have vanishing coefficients in \eqn{2loopNeq4ym}, 
the duality~(\ref{jacobi}) requires the equality of the planar and nonplanar
numerator factors, $n^{\P}_{1234} = n^{\NP}_{1234}$.
Similarly, applying a Jacobi identity to the red propagator marked $b$
in the planar double-box diagram in~\fig{TwoloopIntegralsFigure},
we find two graphs, one of which again contains a vanishing
triangle subgraph.
Therefore the numerator of the planar box graph should be symmetric
under the exchange of legs 1 and 2, or equivalently
$n^{\P}_{1234} = n^{\P}_{3421}$.  Looking at~\eqn{Neq4ymNums},
we see that these two conditions are satisfied.

Having verified that \eqn{2loopNeq4ym} satisfies the BCJ relations,
we may combine two copies of (\ref{2loopNeq4ym}) following
prescription~(\ref{bcj_gravity}) to obtain the two-loop four-graviton
$\NeqEight$ amplitude.  We obtain
\begin{eqnarray}\label{2loopNeq8ym}
{\cal M}^{\twoloop}_4(1,2,3,4) &=& -i \Big( \frac{\ka}{2}\Big)^6 
[st  \, A_4^{\rm tree}(1,2,3,4)]^2 \Bigl(
     \, s^2 \, \I_4^{\P}(s, t)
  +  \, s^2 \, \I_4^{\P}(s, u) \nn \\
&& \null \hskip  3 truecm
  +  \, s^2\, \I_4^{\NP}(s,t)
  +  \, s^2\, \I_4^{\NP}(s,u)
+  {\rm cyclic} \Bigr)\,,
\end{eqnarray}
which is precisely the known result~\cite{TwoLoop}. 
We also recall that the four-graviton and four-gluon tree-level
partial amplitudes are related to each other by
\be
stu M_4^\tree = - i \, [st  \, A_4^{\rm tree}(1,2,3,4)]^2 \,. 
\label{gravYMtree4pt}
\ee
%


\section{Two-Loop $4 \le {\cal N} < 8$ Supergravity}
\label{Neq456masterformula}


  \begin{table*}[h] 
  \begin{center}
    \begin{tabular} {| c || c | c | c | c | c | }
      \hline
     helicity & 0 & $+1/2$ & $+1$ & $+3/2$ & $+2$  \\ \hline \hline
     $\NeqEight$ supergravity & 70 & 56 & 28 & 8 & 1 \\ \hline
     $\NeqSix$ supergravity & 30 & 26 & 16 & 6 & 1 \\ \hline
     $\NeqFive$ supergravity & 10 & 11 & 10 & 5 & 1 \\ \hline
     $\NeqFour$ supergravity & 2 & 4 & 6 & 4 & 1 \\ \hline \hline
     $\NeqFour$ sYM & 6 & 4 & 1 &&\\ \hline
     $\NeqTwo$ sYM & 2 & 2 & 1 && \\ \hline
     $\NeqOne$ sYM & & 1 & 1 && \\ \hline
     ${\cal N} = 0$ sYM & & & 1 && \\ \hline
    \end{tabular}
\caption{State multiplicity as a function of helicity for relevant
supersymmetric multiplets in pure supergravities and super-Yang-Mills
theories.  By CPT invariance, the multiplicity for helicity $-h$ is the
same as that shown for $h$. \label{MultiplicityTable} }
    \end{center}
  \end{table*}


Now we move to the main subject of this paper, the construction of
the two-loop four-graviton amplitudes for ${\cal N}=4,5,6$ supergravity.
As we mentioned earlier, only one of the two gauge-theory amplitudes
entering the double-copy formula~(\ref{bcj_gravity}) needs to satisfy the
BCJ duality. We will combine the duality-satisfying $\NeqFour$ sYM
amplitude~(\ref{2loopNeq4ym}) with four-gluon amplitudes for
${\cal N} \equiv \NYM = 0,1,2$ sYM, in order to obtain
the corresponding two-loop four-graviton amplitudes in supergravities
with ${\cal N} = 4 + \NYM = 4,5,6$.
Looking at the multiplicities of states for various supergravities
and super-Yang-Mills theories in \tab{MultiplicityTable}, we can see that
at the level of counting states,
\begin{eqnarray}
\NeqSix \ \rm{supergravity} &:& 
\quad (\NeqFour \ \rm{sYM} ) \times (\NeqTwo \ \rm{sYM}) \,, \nn \\
\NeqFive \ \rm{supergravity} &:&
\quad (\NeqFour \ \rm{sYM} ) \times (\NeqOne \ \rm{sYM}) \,, \nn \\
\NeqFour \ \rm{supergravity} &:&
\quad (\NeqFour \ \rm{sYM} ) \times (\NeqZero \ \rm{sYM}) \,,
\end{eqnarray}
where $\NeqZero$ sYM refers to pure Yang-Mills theory with only gluons.
Because the gauge theories with ${\cal N} < 4$ supersymmetry are
consistent truncations of maximal $\NeqFour$ sYM, and similarly
on the gravity side, these equivalences also hold at the level of amplitudes,
through either the KLT relations (at tree level) or the double-copy
relations~(\ref{bcj_gravity}).

In ref.~\cite{DDDM}, it was shown that one could write a color
decomposition of any one-loop full-color all-adjoint gauge-theory
amplitude in terms of color factors called ``ring diagrams''.  The
diagrammatic representation of these color factors
have all the external legs connected directly to the loop.  Other
conceivable color factors, in which nontrivial trees are attached to
the loop, can be removed systematically by using Jacobi identities,
in favor of ring graphs with different cyclic orderings of the external
legs.  This decomposition is independent of the (adjoint) particle
content in the loop.  In the same way, we can use the Jacobi identities at
two loops to rewrite any full-color four-gluon amplitude in a theory
with only adjoint particles, in terms of only the color factors
$c^{\P}_{1234}$ and $c^{\NP}_{1234}$ of the diagrams
of~\fig{TwoloopIntegralsFigure} (plus permutations).

For super-Yang-Mills theory with ${\cal N} = \NYM$ supersymmetries,
we write
\begin{eqnarray} \label{colorBasis}
 && \hskip -0.9truecm 
\mathcal{A}^{\twoloop}_{\NYM}(1,2,3,4) \nn\\
 && \hskip 0truecm 
 = -g^6 \Big( c^{\P}_{1234}\, A^{\P}_{1234, \,\NYM}
            + c^{\P}_{3421}\, A^{\P}_{3421, \,\NYM}
           + c^{\NP}_{1234}\, A^{\NP}_{1234, \,\NYM}
           + c^{\NP}_{3421}\, A^{\NP}_{3421, \,\NYM} \nn \\
&& \hskip 1truecm
         + \, c^{\P}_{1342}\, A^{\P}_{1342, \,\NYM}
            + c^{\P}_{4231}\, A^{\P}_{4231, \,\NYM}
           + c^{\NP}_{1342}\, A^{\NP}_{1342, \,\NYM}
           + c^{\NP}_{4231}\, A^{\NP}_{4231, \,\NYM} \nn \\
&& \hskip 1truecm
          + \, c^{\P}_{1423}\, A^{\P}_{1423, \,\NYM}
             + c^{\P}_{2341}\, A^{\P}_{2341, \,\NYM}
            + c^{\NP}_{1423}\, A^{\NP}_{1423, \,\NYM}
            + c^{\NP}_{2341}\, A^{\NP}_{2341, \,\NYM} \Big) \,,
\end{eqnarray}
where $A^{\P}_{1234}$ is the integrated color-ordered subamplitude
associated with the color factor $c^{\P}_{1234}$. For example, for
the $\NeqFour$ sYM representation~(\ref{2loopNeq4ym}), we read off
\begin{equation*}
A^{\P}_{1234, \, \NYM=4}
 =  s t  \, A_4^{\rm tree}(1,2,3,4) \times s  \, \I_4^{\P}(s, t)\,.
\label{APNeq4}
\end{equation*}

Normally, to implement the double-copy formula~(\ref{bcj_gravity}),
we would need to have a representation for the {\it integrand}
of the gauge-theory amplitudes, in particular for the ${\cal N}=0,1,2$
sYM amplitudes we are combining with those for $\NeqFour$ sYM.
However, at two loops the numerator factors for $\NeqFour$ sYM have no
dependence on the loop momenta.  The same feature holds for the
one-loop four- and five-point amplitudes studied in ref.~\cite{Nle8BCJ}.
Therefore, just as in those cases, we can remove the $\NeqFour$ sYM
numerator factors from the loop integrals in \eqn{bcj_gravity}.
Using \eqn{Neq4ymNums} for the $\NeqFour$ sYM numerator factors,
we obtain the remarkably simple general formula,
\begin{eqnarray}
 &&\mathcal{M}^{\twoloop}_{\NYM + 4} (1,2,3,4)
= -i \Big( \frac{\ka}{2} \Big)^6  st  \,
              A_4^{\rm tree}(1,2,3,4) \nn\\
&& \hskip 1truecm \times \Big(
   s \, A^{\P}_{1234, \,\NYM} + s \, A^{\P}_{3421, \,\NYM}
 + s\, A^{\NP}_{1234, \,\NYM} + s\, A^{\NP}_{3421, \,\NYM} \nn \\
&& \hskip 1truecm 
 + \, u \, A^{\P}_{1342, \,\NYM} + u \, A^{\P}_{4231, \,\NYM}
 + u\, A^{\NP}_{1342, \,\NYM} + u\, A^{\NP}_{4231, \,\NYM} \nn \\
&& \hskip 1truecm 
 + \, t\, A^{\P}_{1423, \,\NYM} + t\, A^{\P}_{2341, \,\NYM}
 + t\, A^{\NP}_{1423, \,\NYM} + t\, A^{\NP}_{2341, \,\NYM} \Big) \,.
\label{M42lNeq456}
\end{eqnarray}
In summary, we obtain the $\NeqFour,5,6$ supergravity amplitudes by
first expressing the $\NeqZero,1,2$ sYM helicity amplitudes from
ref.~\cite{TwoLoopYMn2} in terms of the color
basis~(\ref{colorBasis})\footnote{%
We thank Zvi Bern for providing us with
the expressions in this format.}.
We then replace $g^6 \to i(\ka/2)^6$ and perform the following
additional replacements (plus their relabelings):
\begin{equation}
c^{\P}_{1234} \rightarrow s t A_4^{\rm tree}(1,2,3,4) \times s, 
\hskip 2truecm
c^{\NP}_{1234} \rightarrow s t A_4^{\rm tree}(1,2,3,4)\times s.
\end{equation} 
Because $s t A_4^{\rm tree}(1,2,3,4)$ is permutation-invariant,
only the single factors of $s,t,u$ persist inside the parentheses
in \eqn{M42lNeq456}.

In order to preserve supersymmetry, we use the four-dimensional helicity
variant of dimensional regularization~\cite{FDH} for both copies of the
gauge-theory amplitudes.  The results~(\ref{M42lNeq456}) can be expressed
in terms of master integrals for the two-loop planar and nonplanar
double-box topologies, plus various other integrals with fewer propagators
present.  However, in this form the results are rather lengthy.  Instead
of presenting them here, we expand the dimensionally-regulated results,
for $D=4-2\e$, around $\e=0$, as discussed in the next section.


\section{Infrared poles and finite remainders}
\label{IRandFiniteSection}

At two loops, all pure supergravity amplitudes are ultraviolet
finite~\cite{Grisaru,DKS,Tomboulis}.  Therefore all of their divergences
are infrared in nature, either soft or possibly collinear. 
As two massless external particles become collinear, gravitational
tree amplitudes have singularities only in phase, not in magnitude.
The same universal ``splitting amplitude'' that controls the phase
behavior governs loop amplitudes as well as tree
amplitudes~\cite{Multileg}.  Correspondingly, there are no
virtual divergences from purely collinear regions of
integration~\cite{StermanGrav}.
Soft divergences were studied long ago and found to
exponentiate~\cite{Weinberg1965nx}.  More recent, explicit
analyses can be found in
refs.~\cite{White2011yy,Naculich2011ry,StermanGrav}.
At one loop, the IR pole behavior
is~\cite{Weinberg1965nx,DunbarNorridge,Naculich2008ew,BHNST2008},
\be
{\cal M}_4^{\oneloop} = \Bigl(\frac{\ka}{8\pi}\Bigr)^2
\, \frac{2}{\e} \, \Bigl( s \, \ln(-s) + t \, \ln(-t) + u \, \ln(-u) \Bigr)
\, {\cal M}_4^{\tree}\ +\ \Ord(\e^0).
\label{oneloopIR}
\ee
At $L$ loops, the leading divergence is at order $1/\e^L$.
We first checked that the leading divergence of our two-loop
$\NeqFour,5,6$ supergravity amplitudes is indeed at order $1/\e^2$.

Moreover, the exponentiation of soft divergences implies that the
full two-loop IR behavior can be expressed in terms of the one-loop
amplitude as follows:
\begin{equation}\label{IRexp}
\frac{ {\cal M}_4^{\twoloop}(\e) }{ {\cal M}_4^{\tree} }
= \frac{1}{2} 
\biggl[ \frac{ {\cal M}_4^{\oneloop} (\e) }
             { {\cal M}_4^{\text{tree}} } \biggr]^2
+\Bigl(\frac{\ka}{8\pi}\Bigr)^4 \, F_4^{\twoloop} \, + \, \Ord(\e) \,,
\end{equation}
where $F_4^{\twoloop}$ is the finite remainder in the limit $\e\to0$.
This infrared behavior was checked explicitly for the four-point
$\NeqEight$~supergravity amplitude~\cite{Naculich2008ew,BHNST2008}, and
was conjectured to hold for all supersymmetric gravity
amplitudes~\cite{Naculich2011ry}.  We have checked that our expressions
indeed satisfy~\eqn{IRexp}.  We remark that the lack of any additional
(ultraviolet) poles in $\e$ confirms the absence of UV divergences for
${\cal N}=4,5,6$ supergravity in four dimensions at two
loops~\cite{Grisaru,DKS,Tomboulis}.

In order to verify~\eqn{IRexp} and extract $F_4^{\twoloop}$,
we need the $\Ord(\e^0)$ and $\Ord(\e^1)$ coefficients in the expansion
of the corresponding one-loop amplitude ${\cal M}_4^{\oneloop}$.
That is because ${\cal M}_4^{\oneloop}$ appears squared in \eqn{IRexp},
and the $1/\e$ pole in \eqn{oneloopIR} can multiply the $\Ord(\e^1)$
coefficient to generate a finite term.  We give the required one-loop
expansions in \app{OneLoopAppendix}.

Next we present the finite remainders $F_4^{\twoloop}$ for
the different theories under consideration.
It is convenient to express the remainders for ${\cal N}<8$ supergravity
in terms of the $\NeqEight$ remainder plus an additional term.
The result for $\NeqEight$ supergravity
was first presented in refs.~\cite{Naculich2008ew,BHNST2008}.
We always consider the helicity configuration $(1^-,2^-,3^+,4^+)$.
There are three separate physical kinematic regions:
the $s$ channel, with $s>0$ and $t,u<0$; 
the $t$ channel ($t>0$ and $s,u<0$); and
the $u$ channel ($u>0$ and $s,t<0$).
The $s$ channel is singled out by the fact that it has identical-helicity
incoming gravitons.
For all the supergravity theories, the $(1^-,2^-,3^+,4^+)$ helicity
configuration chosen is symmetric under $3 \lr 4$.  Therefore we do not have
to present results separately for the $u$ channel; they can be obtained
from the $t$-channel results by relabeling $t \lr u$.
In the case of $\NeqEight$ supergravity, an $\NeqEight$ supersymmetric
Ward identity implies that the results in the $t$
channel (normalized by the tree amplitude) can be obtained simply by
relabeling $s\lr t$.  For ${\cal N}<8$,
this property no longer holds, and we will have to quote the $s$- and
$t$-channel results separately.

The $\NeqEight$ finite remainder
was expressed in refs.~\cite{Naculich2008ew,BHNST2008} partly in terms
of Nielsen polylogarithms $S_{n,p}(x)$. Here we give a representation similar
to ref.~\cite{Naculich2008ew}, and a second representation entirely in
terms of classical polylogarithms ${\rm Li}_n$, for consistency with the
forms we present below for ${\cal N}<8$.  The finite remainder is
\bea
F_4^{\twoloop,\,\NeqEight} \Big|_{s-{\rm channel}}
&=& 8 \, \biggl\{ t \, u \, \Bigl[ f_1\Bigl(\frac{-t}{s}\Bigr)
                                 + f_1\Bigl(\frac{-u}{s}\Bigr) \Bigr]
                + s \, u \, \Bigl[ f_2\Bigl(\frac{-t}{s}\Bigr)
                                 + f_3\Bigl(\frac{-t}{s}\Bigr) \Bigr]
\nonumber\\
&&\hskip 0.5cm
                + s \, t \, \Bigl[ f_2\Bigl(\frac{-u}{s}\Bigr)
                                 + f_3\Bigl(\frac{-u}{s}\Bigr) \Bigr]             
     \biggr\} \,,
\label{NeqEight2lfinites}
\eea
where
\bea
f_1(x) &=& S_{1,3}(1-x) + \zeta_4 + \frac{1}{24} \ln^4 x
+ i\pi \Bigl[ - S_{1,2}(1-x) + \zeta_3 + \frac{1}{6} \ln^3 x \Bigr]
\nonumber\\
&=& \null
- \polylog_4(x) + \ln x \, \polylog_3(x)
- \frac{1}{2} \, \ln^2 x \, \polylog_2(x)
+ \frac{1}{24} \, \ln^4 x
- \frac{1}{6} \, \ln^3 x \, \ln(1-x)
+ 2 \, \zeta_4
\nonumber\\ &&\null\hskip0cm
+ i\pi \, \biggl[ \polylog_3(x) - \ln x \, \polylog_2(x)
         + \frac{1}{6} \, \ln^3 x
         - \frac{1}{2} \, \ln^2 x \, \ln(1-x) \biggr] \,,
\label{f1NeqEight}
\eea
\bea
f_2(x) &=& S_{1,3}\left(1-\frac{1}{x}\right) + \zeta_4 
 + \frac{1}{24} \, \ln^4 x
 + i \pi \, \left[ S_{1,2}\left(1-\frac{1}{x}\right)
                    - \zeta_3 + \frac{1}{6} \, \ln^3 x \right]
\nonumber\\
&=& 
  \polylog_4(x) - \ln x \, \polylog_3(x) 
+ \frac{1}{2} \, \ln^2 x \, \polylog_2(x)
+ \frac{1}{6} \, \ln^3 x \, \ln(1-x) 
\nonumber\\ &&\null\hskip0cm
- i \pi \, \Bigl[ \polylog_3(x) - \ln x \, \polylog_2(x)
       - \frac{1}{2} \, \ln^2 x \, \ln(1-x) \Bigr] \,,
\label{f2NeqEight}
\eea
and
\bea
f_3(x) &=& \polylog_4(y) - \ln(-y) \, \polylog_3(y)
+ \frac{1}{2} \, \Bigl[ \ln^2(-y) + \pi^2 \Bigr] \polylog_2(y)
\nonumber\\ &&\null\hskip0cm
+ \frac{1}{6} \, \Bigl[ \ln^3(-y)  + 3 \, \pi^2 \, \ln(-y)
                      - 2 \, i \, \pi^3 \Bigr] \, \ln(1-y) \,,
\label{f3NeqEight}
\eea
with $y=-x/(1-x)$.
The $\NeqEight$ supergravity remainder in the $t$ channel
is given simply by relabeling the $s$-channel result,
exchanging $s$ and $t$:
\be
F_4^{\twoloop,\,\NeqEight}(s,t,u) \Big|_{t-{\rm channel}}
= F_4^{\twoloop,\,\NeqEight}(t,s,u) \Big|_{s-{\rm channel}} \,.
\label{NeqEight2lfinitet}
\ee

It was noted previously~\cite{Naculich2008ew,BHNST2008} that
$F_4^{\twoloop,\,\NeqEight}$ has a uniform maximal transcendentality.
That is, all functions appearing are degree-four combinations of
polylogarithms, logarithms, and transcendental constants.
A {\it pure function} is a function with a uniform degree of
transcendentality, having only constants (rational numbers)
multiplying the combinations of polylogarithms, {\it etc.}
A pure function $f$ has a well-defined {\it symbol},
${\cal S}(f)$, which can be obtained by an iterated
differentiation procedure~\cite{symbolsC,symbolsB,symbols,Goncharov2010jf}.
In the representation~(\ref{NeqEight2lfinites}),
the functions $f_1$, $f_2$ and $f_3$ are pure functions with very
simple, one-term symbols:
\bea
{\cal S}(f_1) &=& x \otimes x \otimes x \otimes \frac{x}{1-x} \,,
\label{symbolf1} \\
{\cal S}(f_2) &=& x \otimes x \otimes x \otimes (1-x) \,, 
\label{symbolf2} \\
{\cal S}(f_3) &=&
-\ \ \frac{x}{1-x} \otimes \frac{x}{1-x} \otimes \frac{x}{1-x} \otimes (1-x)
 \,. \label{symbolf3}
\eea
We have shuffled terms slightly with respect to
refs.~\cite{Naculich2008ew,BHNST2008} in order to make this property
manifest.  For example, our function $f_1(x)$ is very similar to the
function $h(t,s,u)$ given in eq.~(2.26) of ref.~\cite{Naculich2008ew},
after multiplying it by $1/8$ and setting $-s/t \to x$.  However,
\eqn{f1NeqEight} contains a term $\tfrac{1}{24} \ln^4 x$ in place
of the term $\tfrac{1}{24} \ln^4 (1-x)$ in $h/8$.  Because only the
sum $f_1(x) + f_1(1-x)$ appears in \eqn{NeqEight2lfinites}, this
swap of terms does not affect the total, but it does ensure that the
branch cut origins are in the same place for all terms in $f_1$, and
correspondingly it simplifies the symbol ${\cal S}(f_1)$.  The functions
$f_2$ and $f_3$ are related to $f_1$
by crossing: $f_2$ by the map $x \to 1/x$ ($s\lr t$),
and $f_3$ by the map $x \to -(1-x)/x$ ($s\to t\to u\to s$).

Curiously, the symbol of $f_1$ obeys a certain
``final entry'' condition recently observed to appear in the context
of the remainder function for planar $\NeqFour$ sYM amplitudes or Wilson
loops~\cite{CaronHuot2011ky,Dixon2011pw}.  Furthermore,
$f_1(x)$ obeys the generalization of this condition to functions, namely
\be
\frac{df_1}{dx} = \frac{p(x)}{x(1-x)} \,,
\label{ultrapure}
\ee
where $p(x)$ is also a pure function, in this case
\be
p(x) = \frac{1}{6} \ln^3 x + \frac{i\pi}{2} \ln^2 x \,.
\label{purep}
\ee
When the finite remainder of the four-graviton amplitude
in $\NeqEight$ supergravity becomes available at three loops
(for example by computing the integrals for one of the three available
expressions for it~\cite{ThreeLoop,BCJLoop}), it will
be very interesting to see whether it can also be expressed in terms
of pure functions of degree six with simple symbols. Perhaps the functions
will even obey a relation like \eqn{ultrapure}.

We return now to two loops and ${\cal N} < 8$ supergravity.
We present the finite remainder for $\NeqSix$ supergravity,
first in the $s$ channel:
\be
F_4^{\twoloop,\,\NeqSix} \Big|_{s-{\rm channel}}
= F_4^{\twoloop,\,\NeqEight} \Big|_{s-{\rm channel}}
 + t \, u \, \biggl[ f_{6,s}\Bigl(\frac{-t}{s}\Bigr)
                   + f_{6,s}\Bigl(\frac{-u}{s}\Bigr) \biggr] \,,
\label{NeqSix2lfinites}
\ee
where
\be
f_{6,s}(x) = f_{6,s;4}(x) + f_{6,s;3}(x)
\label{f6ssplit}
\ee
gives the decomposition into a degree-four function,
\bea
f_{6,s;4}(x) &=& 
20 \, \polylog_4(x)
- 4 \, (1-x) \, \polylog_4\left(\frac{-x}{1-x}\right)
- 12 \, \ln x \, \polylog_3(x)
  + 4 \, \ln^2 x \, \polylog_2(x)
\nonumber\\ &&\null\hskip0cm
  - 4 \, (1-x) \, \ln\left(\frac{x}{1-x}\right) \, \polylog_3(x)
  - \frac{1}{4} \, x\,(1-x) 
     \, \biggl[ \ln^4\biggl(\frac{x}{1-x}\biggr) + \pi^4 \biggr]
\nonumber\\ &&\null\hskip0cm
  + \frac{\pi^2}{2} \, \Bigl[ x \, \ln x + (1-x) \, \ln(1-x) \Bigr]^2
  + \frac{2}{3} \, x \, \ln^4 x
\nonumber\\ &&\null\hskip0cm
  - \frac{2}{3} \, \ln x \, \ln(1-x)
     \, \Bigl[ (1+x) \, \ln^2 x - \frac{9}{4} \, \ln x \, \ln(1-x) \Bigr]
\nonumber\\ &&\null\hskip0cm
 - 4 \, \zeta_2 \, \Bigl[ x \, \polylog_2(x) + 2 \, \ln x \, \ln(1-x) \Bigr]
  - \frac{41}{2} \, \zeta_4
\nonumber\\ &&\null\hskip0cm
  + i \, \pi \, \biggl[ - 12 \, \polylog_3(x)
      + 8 \, \ln x \, ( \polylog_2(x) + \zeta_2 )
      - \frac{2}{3} \, (1-2\,x) \, \ln x \, ( \ln^2 x + \pi^2 )
\nonumber\\ &&\null\hskip1cm
      + 4 \, (1-x) \, \ln^2 x \, \ln(1-x) \biggr] \,,
\label{f6s4}
\eea
and a degree-three one,
\bea
f_{6,s;3}(x) &=& 
 - \frac{4}{3} \, x \, \ln x
   \, \Bigl[ \ln^2 x + 3 \, \ln^2(1-x) + \pi^2 \Bigr]
\nonumber\\ &&\null\hskip0cm
  + 8 \, \Bigl[ \polylog_3(x) - \ln x \, \polylog_2(x)
        - \frac{\zeta_3}{2} + i\pi \, \zeta_2 \Bigr] \,.
\label{f6s3}
\eea
It has been observed~\cite{BHNST2008} that at one loop the
four-graviton amplitude in $\NeqSix$ supergravity has maximal
transcendentality (degree two).  This result extends to one-loop
amplitudes with more gravitons, thanks to the absence of bubble
integrals~\cite{DunbarFivePts,Nle8BCJ}.
However, the degree-three nature of \eqn{f6s3} shows that this
property is broken at two loops.  The breaking comes from both the
two-loop amplitude ${\cal M}_4^{\twoloop}$, but also from the square
of the one-loop amplitude ${\cal M}_4^{\oneloop}$, which has
to be subtracted in \eqn{IRexp}.  As can be seen from \eqns{Neq61ls}{Neq61lt},
the one-loop $\NeqSix$ amplitude has 
degree-two terms as well as degree-three terms at $\Ord(\e)$; the former
terms multiply the $1/\e$ degree-one terms from the IR pole shown in
\eqn{oneloopIR} to generate degree-three contributions to \eqn{f6s3}.
On the other hand, these contributions are purely logarithmic; the
polylogarithmic terms in \eqn{f6s3} can be traced to
${\cal M}_4^{\twoloop}$.  The complexity of the
expressions~(\ref{f6s4}) and (\ref{f6s3}), in terms of their power-law
dependence on $x$, makes it unprofitable to try to separate
the ${\cal N} < 8$ finite remainders into pure functions and to compute
their symbols.

Because of the helicity assignment $(1^-,2^-,3^+,4^+)$, the $s$-channel
remainder is always symmetric under $t \lr u$.  However, in the
$t$ channel there is no such symmetry.  The $\NeqSix$ remainder
in this channel is,
\be
F_4^{\twoloop,\,\NeqSix} \Big|_{t-{\rm channel}}
= F_4^{\twoloop,\,\NeqEight} \Big|_{t-{\rm channel}}
 + t \, u \, \biggl[ f_{6,t;4}\Bigl(\frac{-u}{t}\Bigr)
                   + f_{6,t;3}\Bigl(\frac{-u}{t}\Bigr) \biggr] \,,
\label{NeqSix2lfinitet}
\ee
where the degree-four part is
\bea
f_{6,t;4}(x) &=& 
- 20 \, \polylog_4(1-x) - 20 \, \polylog_4\left(\frac{1-x}{-x}\right)
- 4 \, \frac{1+x}{1-x} \, \Bigl( \polylog_4(x) - \zeta_4 \Bigr)
+ 16 \, \ln x \, \polylog_3(1-x)
\nonumber\\ &&\null\hskip0cm
- 12 \, \ln(1-x) \, \Bigl( \polylog_3(x) - \zeta_3 \Bigr)
+ 4 \, \frac{4-3\,x}{1-x} \, \ln x
    \, \Bigl[ \polylog_3(x) - \zeta_3
             + \frac{1}{2} \, \ln(1-x) \, \ln^2 x \Bigr]
\nonumber\\ &&\null\hskip0cm
+ 4 \, \ln x \, \Bigl( \ln x - 2 \, \ln(1-x) \Bigr) \, \polylog_2(1-x)
+ 4 \, \zeta_2 \, \frac{7-5\,x}{1-x} \, \polylog_2(1-x)
\nonumber\\ &&\null\hskip0cm
- \frac{1}{6} \, \frac{5-8\,x}{(1-x)^2} \, \ln^4 x
- 6 \, \ln^2(1-x) \, \ln^2 x
- 2 \, \zeta_2 \, \frac{13-19\,x+12\,x^2}{(1-x)^2} \, \ln^2 x
\nonumber\\ &&\null\hskip0cm
+ 16 \, \zeta_2 \, \frac{1-2\,x}{1-x} \, \ln x \, \ln(1-x)
\ +\ i \, \pi \, \biggl[ 16 \, \polylog_3(1-x)
           + \frac{4}{1-x} \, \Bigl( \polylog_3(x) - \zeta_3 \Bigr)
\nonumber\\ &&\null\hskip1cm
           - 8 \, \ln(1-x) \, \polylog_2(1-x)
           + \frac{2}{3} \, \frac{1-2\,x+4\,x^2}{(1-x)^2} \, \ln^3 x
           + 2 \, \frac{2+x}{1-x} \, \ln^2 x \, \ln(1-x)
\nonumber\\ &&\null\hskip1cm
           - 2 \, \ln x \, \ln^2(1-x)
           - 4 \, \zeta_2 \, \frac{4-x}{1-x} \, \ln x \biggr] \,,
\label{f6t4}
\eea
and the degree-three part is
\bea
f_{6,t;3}(x) &=& 
 \frac{4}{3} \, \frac{x}{1-x} \, \ln x \, \Bigl( \ln^2 x - 2 \, \pi^2 \Bigr)
   - 8 \, \Bigl( \polylog_3(x) - \ln x \, \polylog_2(x) \Bigr) 
\nonumber\\ &&\null\hskip0cm
   + 4 \, \ln(1-x) \, \Bigl( \ln^2 x - 4 \, \zeta_2 \Bigr)
   + 4 \, i \, \pi \, \biggl[ \frac{x}{1-x} \, \ln^2 x
           - 2 \, \Bigl( \polylog_2(1-x) + \zeta_2 \Bigr) \biggr] \,.
\nonumber\\ &&\null\hskip0cm {~}
\label{f6t3}
\eea

In the $s$ channel, the finite remainder for $\NeqFive$ supergravity
at two loops is given by,
\be
F_4^{\twoloop,\,\NeqFive} \Big|_{s-{\rm channel}}
= F_4^{\twoloop,\,\NeqEight} \Big|_{s-{\rm channel}}
 + t \, u \, \biggl[ f_{5,s}\Bigl(\frac{-t}{s}\Bigr)
                   + f_{5,s}\Bigl(\frac{-u}{s}\Bigr) \biggr] \,,
\label{NeqFive2lfinites}
\ee
where
\be
f_{5,s}(x) = f_{5,s;4}(x) + f_{5,s;3}(x) + f_{5,s;2}(x)
\label{f5ssplit}
\ee
gives the decomposition into a degree-four function,
\bea
f_{5,s;4}(x) &=& 
- 12 \, \biggl\{ (1-x) \, \biggl[ 
         \polylog_4\left(\frac{-x}{1-x}\right)
          - \zeta_2 \, \polylog_2(x) \biggr]
       - 2 \, \Bigl( 1 + x\,(1-x) \Bigr) \, \polylog_4(x)
\nonumber\\ &&\null\hskip1cm
       + \Bigl[ (2-x^2) \, \ln x - (1-x)^2 \, \ln(1-x) \Bigr]
             \, \polylog_3(x)
       - \frac{1}{2} \, \ln^2 x \, \polylog_2(x) \biggr\}
\nonumber\\ &&\null\hskip0cm
- \frac{1}{16} \, x\,(1-x) 
     \, \biggl[ 5 \, \ln^4\left(\frac{x}{1-x}\right)
         + 34 \, \pi^2 \, \ln^2\left(\frac{x}{1-x}\right) \biggr]
+ \frac{1}{2} \, x \, \ln^4 x
\nonumber\\ &&\null\hskip0cm
- (1-x) \, \ln^3 x \, \ln(1-x)
+ \frac{3}{4} \, \Bigl( 3 - 4\,x\,(1-x) \Bigr) \, \ln^2 x \, \ln^2(1-x)
\nonumber\\ &&\null\hskip0cm
+ \frac{\pi^2}{2} \, \biggl[ -(1-x)\,(3-2\,x) \, \ln^2 x
                       + \frac{3}{2} \, \ln^2\left(\frac{x}{1-x}\right) 
                     \biggr]
- \frac{3}{8} \, \zeta_4 \, \Bigl( 72 + 323 \, x\,(1-x) \Bigr)
\nonumber\\ &&\null\hskip0cm
+ i \, \pi \, \biggl\{
- 12 \, \Bigl[ \Bigl( 1 + 2\,x\,(1-x) \Bigr) \, \polylog_3(x)
      - \ln x \, \polylog_2(x) \Bigr]
\nonumber\\ &&\null\hskip1cm
- (1-2\,x)\,(1-x) \, \ln x \, \Bigl( \ln^2 x + \pi^2 \Bigr)
+ 3 \, \Bigl( 2 \, (1-x)^2 + x \Bigr) \, \ln^2 x \, \ln(1-x)
\nonumber\\ &&\null\hskip1cm
+ 2 \, \pi^2 \, \ln x \biggr\}
\,,
\label{f5s4}
\eea
a degree-three function,
\bea
f_{5,s;3}(x) &=& 
12 \, \biggl\{ (1+x^2) \, \Bigl[ \polylog_3(x)
                       - \ln x \, \polylog_2(x) \Bigr]
        - \frac{1}{2} \, \Bigl( 1 - x\,(1-x) \Bigr) \, \ln^2 x \, \ln(1-x)
      \biggr\}
\nonumber\\ &&\null\hskip0cm
        - 2 \, x\,(1-x) \, \ln^3 x
        - 4 \, \pi^2 \, x \, \ln x
        - 12 \, \zeta_3
\ +\ 12 \, \pi \, i \, \biggl[ (1-x) \, \polylog_2(x)
\nonumber\\ &&\null\hskip1cm
       + \frac{1}{4} \, \Bigl( x \, \ln x + (1-x) \, \ln(1-x) \Bigr)^2
                     + \frac{\zeta_2}{2} \, \bigl( 2 - 3\,x\,(1-x) \Bigr)
                        \biggr] \,,
\label{f5s3}
\eea
and a degree-two function,
\bea
f_{5,s;2}(x) &=& 
- 3 \, \biggl[ \Bigl( x \, \ln x + (1-x) \, \ln(1-x) \Bigr)^2
         - \pi^2 \, x\,(1-x)
         + 4 \, \pi \, i \, x \, \ln x \biggr] \,.
\label{f5s2}
\eea

The $\NeqFive$ remainder function in the $t$ channel is,
\be
F_4^{\twoloop,\,\NeqFive} \Big|_{t-{\rm channel}}
= F_4^{\twoloop,\,\NeqEight} \Big|_{t-{\rm channel}}
 + t \, u \, \biggl[ f_{5,t;4}\Bigl(\frac{-u}{t}\Bigr)
                   + f_{5,t;3}\Bigl(\frac{-u}{t}\Bigr)
                   + f_{5,t;2}\Bigl(\frac{-u}{t}\Bigr)\biggr] \,,
\label{NeqFive2lfinitet}
\ee
where the degree-four part is
\bea
f_{5,t;4}(x) &=& 
12 \, \Biggl\{ - \frac{1+x}{1-x} \, \Bigl( \polylog_4(x) - \zeta_4 \Bigr)
  - 2 \, \biggl( 1 - \frac{x}{(1-x)^2} \biggr)
   \, \biggl[ \polylog_4(1-x) + \polylog_4\left(\frac{1-x}{-x}\right)
      \biggr]
\nonumber\\ &&\null\hskip0.3cm
  - \biggl[ \biggl( 1 - \frac{2\,x}{(1-x)^2} \biggr) \, \ln(1-x)
   - \biggl( 2 - \frac{x^2}{(1-x)^2} \biggr) \, \ln x \biggr]
      \, \Bigl( \polylog_3(x) - \zeta_3 \Bigr)
\nonumber\\ &&\null\hskip0.3cm
  + 2 \, \ln x \, \polylog_3(1-x)
  - \frac{1}{2} \, \ln x \, \Bigl( \ln x - 2 \, \ln(1-x) \Bigr)
                \, \polylog_2(x)
\nonumber\\ &&\null\hskip0.3cm
  + 2 \, \zeta_2 \, \frac{2-x}{1-x} \, \polylog_2(1-x)
  - \frac{8-21\,x}{96 \, (1-x)^2} \, \ln^4 x
  + \frac{(1-2x)(5-x)}{12 \, (1-x)^2} \, \ln^3 x \, \ln(1-x)
\nonumber\\ &&\null\hskip0.3cm
  + \frac{1}{8} \, \biggl( 3 + \frac{4\,x}{(1-x)^2} \biggr) 
               \, \ln^2 x \, \ln^2(1-x)
  - \frac{\zeta_2}{4} \, \frac{10-12\,x+11\,x^2}{(1-x)^2} \, \ln^2 x
\nonumber\\ &&\null\hskip0.3cm
  + \zeta_2 \, \frac{1-5\,x+2\,x^2}{(1-x)^2} \, \ln x \, \ln(1-x)
\ +\ i \, \pi \, \biggl[ \frac{\polylog_3(x) - \zeta_3}{(1-x)^2}
        + 2 \, \polylog_3(1-x)
\nonumber\\ &&\null\hskip1.0cm
        - \ln(1-x) \, \polylog_2(1-x)
        + \frac{3}{8} \, \frac{x^2}{(1-x)^2} \, \ln^3 x
\nonumber\\ &&\null\hskip1.0cm
        + \frac{1}{24} \, \frac{2+x}{1-x}
                \, \ln^2 x \, \Bigl( \ln x + 6 \, \ln(1-x) \Bigr)
           - \frac{1}{4} \, \ln x \, \ln^2(1-x)
\nonumber\\ &&\null\hskip1.0cm
           - \frac{\zeta_2}{2} \, \frac{4-x}{1-x} \, \ln x
   \biggr] \Biggr\} \,,
\label{f5t4}
\eea
the degree-three part is
\bea
f_{5,t;3}(x) &=& 
12 \, \Biggl\{ \frac{1+x}{1-x} 
          \, \Bigl[ \polylog_3(1-x) - \ln(1-x) \, \polylog_2(1-x)
                   - \frac{1}{2} \, \ln x \, \ln^2(1-x) \Bigr]
\nonumber\\ &&\null\hskip0.3cm
   - \biggl( 1 + \frac{x^2}{(1-x)^2} \biggr)
     \, \Bigl( \polylog_3(x) - \ln x \, \polylog_2(x) \Bigr)
   + \frac{x  \, \ln^3 x}{6 \, (1-x)^2}
   + \frac{1}{2} \, \ln^2 x \, \ln(1-x) 
\nonumber\\ &&\null\hskip0.3cm
   - \zeta_2 \, \biggl[ \frac{x\,(1-4x)}{(1-x)^2} \, \ln x + \ln(1-x)
                \biggr]
   - \zeta_3 \, \frac{1-2x}{(1-x)^2}
\nonumber\\ &&\null\hskip0.3cm
   + i \, \pi \, \biggl[ 
         - \frac{1+(1-x)^2}{(1-x)^2} \, \polylog_2(1-x) 
         - \frac{1}{(1-x)^2} \, \ln x \, \ln(1-x)
\nonumber\\ &&\null\hskip1.0cm
         + \frac{1}{2} \, \frac{x}{1-x}
                 \, \Bigl( \ln^2 x + 2 \, \zeta_2 \Bigr)
   \biggr] \Biggr\} \,,
\label{f5t3}
\eea
and the degree-two part is
\bea
f_{5,t;2}(x) &=& 
- 6 \, \biggl[ \Bigl( \ln(1-x) + \frac{x}{1-x} \, \ln x \Bigr)^2
          - \pi^2 \, \frac{x}{1-x}
\nonumber\\ &&\null\hskip0.8cm
         + \frac{2 \, i \, \pi }{1-x}
              \, \Bigl( \ln(1-x) + \frac{x}{1-x} \, \ln x \Bigr) 
       \biggr] \,.
\label{f5t2}
\eea

The results for $\NeqFour$ supergravity are the lengthiest of all.
In the $s$ channel, the finite remainder for $\NeqFour$ supergravity
at two loops is given by,
\be
F_4^{\twoloop,\,\NeqFour} \Big|_{s-{\rm channel}}
= F_4^{\twoloop,\,\NeqEight} \Big|_{s-{\rm channel}}
 + t \, u \, \biggl[ f_{4,s}\Bigl(\frac{-t}{s}\Bigr)
                   + f_{4,s}\Bigl(\frac{-u}{s}\Bigr) \biggr] \,,
\label{NeqFour2lfinites}
\ee
where
\be
f_{4,s}(x) = f_{4,s;4}(x) + f_{4,s;3}(x) + f_{4,s;2}(x)
                          + f_{4,s;1}(x) + f_{4,s;0}(x)
\label{f4ssplit}
\ee
gives the decomposition into a degree-four function,
\bea
f_{4,s;4}(x) &=&
4 \, \Bigl( 9 - 4\,x\,(1-x) \Bigr) \, \polylog_4(x)
- 4 \, (8-2\,x-9\,x^2+8\,x^3) \, \ln x \, \polylog_3(x)
\nonumber\\ &&\null\hskip0.0cm
- 4 \, (1-x) \, (3 + 3\,x - 8\,x^2) \, \biggl[
         \polylog_4\left(\frac{-x}{1-x}\right)
      + \zeta_2 \, \polylog_2\left(\frac{-x}{1-x}\right)
       - \ln(1-x) \, \polylog_3(x) \biggr]
\nonumber\\ &&\null\hskip0.0cm
+ 4 \, \Bigl(2-x\,(1-x)\Bigr) 
     \, \ln x \, ( \ln x + 2 \, i \, \pi ) \, \polylog_2(x)
- 4 \, i \, \pi \, \Bigl(5-2\,x\,(1-x)\Bigr) \, \polylog_3(x)
\nonumber\\ &&\null\hskip0.0cm
+ \frac{1}{6} \, x\,(4-8\,x-5\,x^2+21\,x^3-9\,x^4+3\,x^5)
\nonumber\\ &&\null\hskip0.7cm
 \times \Bigl(
     \ln^4 x - 4 \, \ln^3 x \, \ln(1-x) + 2 \, \pi^2 \, \ln^2 x
     + \frac{\pi^4}{2} \Bigr)
\nonumber\\ &&\null\hskip0.0cm
- \frac{2}{3} \, (2-x\,(1-x)) \, (1-3\,x) \Bigl(
       \ln^2 x \, ( \ln x - 6 i \pi ) \, \ln(1-x)
     + i \pi \, \ln x \, ( \ln^2 x - \pi^2 ) \Bigr)
\nonumber\\ &&\null\hskip0.0cm
+ \frac{2}{3} \, i \, \pi \, x \, \ln x \, \Bigl[ (2-13\,x+8\,x^2) \, \ln^2 x
                          + 3 \, (4+10\,x-5\,x^2) \, \ln x \, \ln(1-x)
\nonumber\\ &&\null\hskip2.0cm
                          + (14\,(1+x^2)-19\,x) \, \pi^2 \Bigr]
\nonumber\\ &&\null\hskip0.0cm
+ \frac{1}{2} \, (2-x\,(1-x)) \, (1-x\,(1-x))^2 \, \ln x \, \ln(1-x)
           \, \Bigl( 3 \, \ln x \, \ln(1-x) - 2 \, \pi^2 \Bigr)
\nonumber\\ &&\null\hskip0.0cm
- 2 \, \zeta_2 \, x \, (8-16\,x+11\,x^2) \, \ln^2 x
- \frac{3}{2} \, \zeta_4 \, (44-17\,x\,(1-x))
\,,
\label{f4s4}
\eea
a degree-three function,
\bea
f_{4,s;3}(x) &=& 
- \biggl( \frac{53}{6} + x^2 \biggr)
    \, \biggl[ \polylog_3\left(\frac{-x}{1-x}\right)
    - \ln\left(\frac{x}{1-x}\right)
    \, \polylog_2\left(\frac{-x}{1-x}\right) \biggr]
\nonumber\\ &&\null\hskip0.0cm
- \frac{1}{18} \, (59-12\,x^2+8\,x^3+54\,x^4+36\,x\,(1-x)^4) 
\nonumber\\ &&\null\hskip0.7cm
    \times \ln x \, \Bigl[ \ln x \, \Bigl( \ln x - 3 \, \ln(1-x) \Bigr)
                + \pi^2 \Bigr]
\nonumber\\ &&\null\hskip0.0cm
- \biggl( \frac{31}{3} - 12 \, x + 10 \, x^2 \biggr) \, \ln^2 x \, \ln(1-x)
\nonumber\\ &&\null\hskip0.0cm
- i \, \pi \, \biggl[ (1-2\,x) \, \ln^2 x
     - 9 \, x\,(1-x)
    \, \Bigl( \ln x \, \ln\left(\frac{x}{1-x}\right) + \frac{\pi^2}{2} \Bigr)
      \biggr]
\nonumber\\ &&\null\hskip0.0cm
+ \frac{\zeta_2}{3} \, (59-156\,x+132\,x^2) \, ( \ln x + i \, \pi )
- 33 \, \zeta_3 \, x\,(1-x)
\,,
\label{f4s3}
\eea
a degree-two function,
\bea
f_{4,s;2}(x) &=& 
- \frac{6-7\,x+4\,x^2}{2\,(1-x)} \, \ln x \, ( \ln x + 2 \, i \, \pi )
\nonumber\\ &&\null\hskip0.0cm
+ \biggl[ 3\,\Bigl(1+x^2\,(1-x)^2\Bigr) - \frac{13}{3} \, x\,(1-x) \biggr]
  \, \biggl[ \ln x \, \ln\left(\frac{x}{1-x}\right)
              + \frac{\pi^2}{2} \biggr]
\nonumber\\ &&\null\hskip0.0cm
         + \frac{1}{3} \, \zeta_2 \, x^2 
             \, \Bigl( 6\,x^2 - (1-x)\,(23-24\,x) \Bigr)
\,,
\label{f4s2}
\eea
a degree-one function,
\be
f_{4,s;1}(x) = 
- \frac{1}{3} \, x \, \Bigl( 4 \, (1-x)^2 - x\,(1-2\,x) \Bigr)
               \, ( \ln x + i \, \pi )
\,,
\label{f4s1}
\ee
and a rational part,
\be
f_{4,s;0}(x) = - \frac{1}{4} \, \Bigl( 2 + x\,(1-x) \Bigr)
\,.
\label{f4s0}
\ee

The $\NeqFour$ remainder function in the $t$ channel is,
\bea
F_4^{\twoloop,\,\NeqFour} \Big|_{t-{\rm channel}}
&=& F_4^{\twoloop,\,\NeqEight} \Big|_{t-{\rm channel}}
 + t \, u \, \biggl[ f_{4,t;4}\Bigl(\frac{-u}{t}\Bigr)
                   + f_{4,t;3}\Bigl(\frac{-u}{t}\Bigr)
                   + f_{4,t;2}\Bigl(\frac{-u}{t}\Bigr)
\nonumber\\ &&\null\hskip3.8cm
                   + f_{4,t;1}\Bigl(\frac{-u}{t}\Bigr)
                   + f_{4,t;0}\Bigl(\frac{-u}{t}\Bigr)\biggr] \,,
\label{NeqFour2lfinitet}
\eea
where the degree-four part is
\bea
f_{4,t;4}(x) &=&
- 4 \, \biggl( 9 + \frac{4\,x}{(1-x)^2} \biggr)
\biggl[ \polylog_4(1-x) + \polylog_4\left(\frac{1-x}{-x}\right)
\nonumber\\ &&\null\hskip3.5cm
     + \Bigl( \ln(1-x) + \frac{i\pi}{2} \Bigr)
      \, \Bigl( \polylog_3(x) - \zeta_3 \Bigr)
     + \zeta_2 \, \polylog_2(1-x) \biggr]
\nonumber\\ &&\null\hskip0.0cm
- 4 \, \frac{1+x}{1-x} \, \biggl( 3 - \frac{8\,x}{(1-x)^2} \biggr)
    \, \biggl[ \polylog_4(x) - \zeta_4
        - \frac{i\pi}{2} \, \Bigl( \polylog_3(x) - \zeta_3 \Bigr) \biggr]
\nonumber\\ &&\null\hskip0.0cm
+ 4 \, \frac{8-22\,x+11\,x^2-5\,x^3}{(1-x)^3}
\nonumber\\ &&\null\hskip0.5cm
 \times \biggl[ \ln x \, \Bigl( \polylog_3(x) - \zeta_3 \Bigr)
      + \zeta_2 \, \Bigl( 2 \, \polylog_2(1-x) + \ln x \, \ln(1-x) \Bigr)
       \biggr]
\nonumber\\ &&\null\hskip0.0cm
+ 4 \, \biggl( 2 + \frac{x}{(1-x)^2} \biggr) \, \biggl\{
      \Bigl( 2 \, \ln(1-x) + 3 \, i \, \pi \Bigr)
               \, ( \polylog_3(x) - \zeta_3 )
\nonumber\\ &&\null\hskip2.0cm
     + \Bigl( \ln^2 x + 4 \, \zeta_2 \Bigr) \, \polylog_2(1-x)
\nonumber\\ &&\null\hskip2.0cm
     + 2 \, ( \ln x + i \, \pi ) \, \Bigl( 2 \, \polylog_3(1-x)
                           - \ln(1-x) \, \polylog_2(1-x) \Bigr)
\nonumber\\ &&\null\hskip2.0cm
     + i \, \pi \biggl[
       \biggl( \frac{1}{6} + \frac{x\,(1-x^2+x^3)}{2\,(1-x)^4} \biggr)
           \, \ln^3 x
         + \frac{2+x}{2\,(1-x)} \, \ln^2 x \, \ln(1-x)
\nonumber\\ &&\null\hskip2.8cm
         - \frac{1}{2} \, \ln x \, \ln^2(1-x)
         - \zeta_2 \, \frac{4-x}{1-x} \, \ln x \biggr] \biggr\}
\nonumber\\ &&\null\hskip0.0cm
- \frac{9-48\,x+104\,x^2-129\,x^3+87\,x^4-26\,x^5}{6\,(1-x)^6}
      \, \ln^2 x \, ( \ln^2 x - 4 \, \pi^2 )
\nonumber\\ &&\null\hskip0.0cm
+ \frac{2}{3} \, \frac{23-52\,x+49\,x^2-17\,x^3}{(1-x)^3}
      \, \ln^3 x \, \ln(1-x)
- \biggl( 11 + \frac{5\,x}{(1-x)^2} \biggr) \, \ln^2 x \, \ln^2(1-x)
\nonumber\\ &&\null\hskip0.0cm
- 4 \, \zeta_2 \, \frac{x\,(14-9\,x\,(1-x))}{(1-x)^3} \, \ln x \, \ln(1-x)
- 2 \, \zeta_2 \, \frac{43-71\,x+100\,x^2-25\,x^3}{(1-x)^3} \, \ln^2 x
\,,
\nonumber\\ &&\null\hskip0.0cm {~}
\label{f4t4}
\eea
the degree-three part is
\bea
f_{4,t;3}(x) &=& 
- \biggl( \frac{56}{3} + \frac{2\,x}{(1-x)^2} \biggr)
      \, \biggl[ \polylog_3(x) - ( \ln x + i \, \pi ) \, \polylog_2(x)
        - \frac{2}{3} \, i \, \pi^3 - \frac{5}{3} \, \pi^2 \, \ln(1-x)
         \biggr]
\nonumber\\ &&\null\hskip0.0cm
+ \frac{x\,(24-15\,x+13\,x^2-32\,x^3+28\,x^4)}{9\,(1-x)^5}
    \, \ln x \, ( \ln x + i \, \pi ) \, ( \ln x + 2 \, i \, \pi )
\nonumber\\ &&\null\hskip0.0cm
+ \biggl( \frac{1}{3} + 2 \, \frac{5 - 4\,x\,(1-x)}{(1-x)^2} \biggr)
      \, \biggl[ \ln(1-x) \, \Bigl( ( \ln x + i \, \pi )^2 
                                  - 3 \, \pi^2 \Bigr)
               - 2 \, i \, \pi^3 \biggr]
\nonumber\\ &&\null\hskip0.0cm
- \frac{2+10\,x-x^2}{(1-x)^2} \, \Bigl[
              \pi^2 \, ( \ln x - 2 \, \ln(1-x) )
               - i \, \pi \, ( \ln^2 x + \pi^2 ) \Bigr]
\nonumber\\ &&\null\hskip0.0cm
- \frac{1+x}{1-x} \, \biggl[ ( \ln x + i \, \pi ) \, \ln(1-x)
                           \, ( \ln(1-x) + 2 \, i \, \pi )
            + \frac{i\pi}{1-x} \, \ln^2 x \biggr]
\nonumber\\ &&\null\hskip0.0cm
+ 66 \, \zeta_3 \, \frac{x}{(1-x)^2}
\,,
\label{f4t3}
\eea
the degree-two part is
\bea
f_{4,t;2}(x) &=&
- \frac{6-5\,x+3\,x^2}{2\,(1-x)} \, \Bigl( \ln^2 x
       - 2 \, \ln(1-x) \, ( \ln x + i \, \pi ) + \pi^2 \Bigr)
\nonumber\\ &&\null\hskip0.0cm
+ \biggl( 3 + \frac{x\,(13\,(1-x)^2+9\,x)}{3\,(1-x)^4} \biggr)
      \, \Bigl( ( \ln x + i \, \pi )^2 + 2 \, \zeta_2 \Bigr)
\nonumber\\ &&\null\hskip0.0cm
+ \frac{3\,(1+x^2)-8\,x}{2\,x} \, \ln(1-x) \, ( \ln(1-x) + 2 \, i \, \pi )
\nonumber\\ &&\null\hskip0.0cm
+ \zeta_2 \, \frac{14\,(1+x^2)-3\,x}{(1-x)^2}
\,,
\label{f4t2}
\eea
the degree-one part is
\be
f_{4,t;1}(x) =
- \frac{1}{3} \,  \biggl[
  \ln\left(\frac{x}{1-x}\right)
  - \frac{1 + x + 4 \, x^2}{(1-x)^3} \, ( \ln x + i \, \pi ) \biggr] \,,
\label{f4t1}
\ee
and the rational part is
\be
f_{4,t;0}(x) =
- \frac{(2-x)(1-2\,x)}{2\,(1-x)^2} \,.
\label{f4t0}
\ee
%


\section{Forward-scattering limit of the amplitudes}
\label{ForwardSection}

We now inspect the behavior of the two-loop supergravity
amplitudes in the limit of
small-angle, forward scattering, {\it i.e.}~small momentum transfer at
fixed center-of-mass energy. 
In particular, we want to verify the contributions from matter exchange,
versus graviton exchange, in the forward-scattering limit.  The results
are sensitive to the helicity configuration, or for fixed helicity
configuration, to which invariant is time-like and which of the two
space-like invariants is becoming small.

We first consider configurations, or channels, for which the associated
tree-level amplitudes have a pole at small momentum transfer.  These
configurations are dominated by the exchange of soft gravitons, and
require helicity conservation along the forward-going graviton line.
(They also require helicity conservation along the
backward-going line, but this second condition follows automatically
from the first one, for the MHV amplitudes that we study.)
To see the helicity conservation explicitly, we rewrite the tree amplitude as,
\begin{equation}
M_4^{\tree} (1^-, 2^-, 3^+, 4^+)
= - i s^2 \, \Bigg( \frac{1}{t} + \frac{1}{u}  \Bigg) 
 \Bigg[ \frac{\spa1.2}{[12]} \frac{[34]}{\spa3.4}   \Bigg]^2 \,,
\label{Mtreetupoles}
\end{equation}
where the quantity in brackets is a pure phase.  Expanding
\eqn{Mtreetupoles} for small $t$ at fixed $s$ in the physical $s$
channel ($s>0$ kinematics), one gets a leading term of $\Ord(s^2/t)$
as $t\to0$.  Because the $s$-channel amplitude is symmetric under $t
\leftrightarrow u$, one could also have taken the small $u$ limit and
gotten a pole-dominated behavior.  However, in the physical $t$
channel, one has to take $u$ small in order to conserve helicity at
both vertices.  Then the leading tree-level behavior is $\Ord(t^2/u)$
as $u\to0$.  In contrast, the limit of small $s$ in the $t$ channel
violates helicity conservation, and the tree amplitude is heavily
power-law suppressed with respect to the dominant pole behavior,
having a leading term of $\Ord(s^3/t^2)$ as $s\to0$.

Interestingly, in the helicity-conserving channels described above,
the two-loop remainders, $F_4^{(2)}$, for $\NeqFour,5,6,8$
supergravity amplitudes are all power-law suppressed. The
forward-scattering leading behavior is thus fully determined by the
square of the one-loop amplitude. Moreover, the dominant one-loop
behavior is the same for all $4 \le \mathcal{N} \le 8$ supergravity
amplitudes. Namely, at one loop as $t \to 0$ in the $s$ channel, we have
\be
\frac{\mathcal{M}_4^{(1)}(\e)}{\mathcal{M}_4^{\tree} }
 =  \Big(\frac{\kappa}{8\pi}\Big)^2 \, (-2 \pi i) \, s \, 
  \biggl[ \frac{1}{\e} + \ln\left(\frac{s}{-t}\right)
   + \frac{\e}{2} \ln^2\left(\frac{s}{-t}\right) \biggr]
  \,  + \, \Ord(\e^2,t) \,,
\label{oneloopforward}
\ee
and at two loops we have
\be
\frac{\mathcal{M}_4^{(2)}(\e)}{\mathcal{M}_4^{\tree}}
 =  \frac{1}{2} \Biggl[ \frac{\mathcal{M}_4^{(2)}(\e)}{\mathcal{M}_4^{\tree}}
                \Biggr]^2 \,  + \, \Ord(\e,t) \,.
\label{twoloopforward}
\ee
Both equations hold for any number of supersymmetries.
We also verified the analogous equations in the limit $u\to0$ in the
physical $t$ channel ($t>0$ kinematics).

As discussed in refs.~\cite{Eikonal}, in the physical $s$ channel
only the $s$-channel ladder and crossed-ladder diagrams 
(shown in \fig{TwoloopIntegralsFigure} with $s$ flowing horizontally)
contribute to the eikonal limit $t\to0$.
The limit is dominated by graviton exchanges because the coupling of a
particle of spin $J$ exchanged in the channel with small momentum transfer
is proportional to $E^{J}$, where $E$ is the center-of-mass energy.
The $s$-channel ladder and crossed-ladder diagrams allow for the maximum number
of attachments of gravitons to a hard line (one with energy of order $E$).
This property explains why \eqns{oneloopforward}{twoloopforward} are
independent of the number of supersymmetries at high energy. 
The possible Reggeization of gravity, discussed in ref.~\cite{ReggeGravity},
remains an open question. However, this issue cannot be resolved by studying
forward-scattering or eikonal limits.  The $t$-channel ladder diagrams
(obtained from \fig{TwoloopIntegralsFigure} by rotating by 90$^\circ$
or permuting $1\to2\to3\to4\to1$), which should contribute to
Reggeization, are subleading by powers of $t/s$
because they have fewer attachments to the high-energy lines.

It is also interesting to consider the helicity-violating limit in
which $s\to0$ for $t>0$ kinematics $(u \simeq -t)$.
As mentioned before, the associated tree-level amplitude
is power-suppressed in this limit with respect to the dominant pole
behavior; its leading behavior is $\Ord(s^3/t^2)$. 
In this limit, many of the finite-remainder expressions naively appear
to blow up (see for instance \eqn{f6t4} as $x \to 1$). However, one can
check in all cases that these spurious singularities cancel, and the leading
behavior of the ratio of the one- and two-loop amplitudes to the tree
amplitude is of $\Ord(t^L)$, $L=1,2$.  Thus the one- and two-loop amplitudes
never have a power ($1/s$) enhancement over the tree amplitude in
the helicity-violating limit, but are of the same order in $s$.
(There is a $\ln(s)$ enhancement, but only in the pure $\NeqEight$
supergravity terms, not in any of the matter contributions.) 


\section{Conclusions}
\label{ConclusionSection}

In this paper, we have computed the full four-graviton two-loop
amplitudes in $\NeqFour, 5, 6$ supergravity. As expected, their IR
divergences can be expressed in terms of the square of the corresponding
one-loop amplitudes.  The finite remainders were presented in a simple
form.  We also noted that the finite remainder in $\NeqEight$ supergravity
can be expressed in terms of permutations of a pure function $f_1(x)$
possessing a simple, one-term symbol.

The $\NeqFour, 5, 6$ supergravity results were obtained using the
double-copy property of gravity, which is a consequence of the
recently-conjectured BCJ duality. The former property allowed us to
combine the BCJ-satisfying $\NeqFour$ sYM representation with known
$\NeqZero, 1, 2$ sYM gauge-theory amplitudes, in order to obtain the
corresponding supergravity amplitudes, including all loop integrations.

Our task was vastly simplified by the fact that both sets of Yang-Mills
amplitudes entering the double-copy formula were known,
as well as by the lack of loop-momentum dependence for the $\NeqFour$
sYM amplitudes in this case.  As mentioned in the introduction, generic
${\cal N} < 8$ supergravity theories are expected to diverge at
three loops (but not ${\cal N}=5$ or 6~\cite{BHS,BHSV}),
because the counterterm $R^4$ is allowed by supersymmetry.
It would thus be very interesting to compute explicit three-loop
non-maximal supergravity amplitudes.  If one computes in 
${\cal N} \geq 4$ supergravity, then one can use the double-copy formula,
because a BCJ-satisfying form exists for one of the two copies,
namely the three-loop $\NeqFour$ sYM amplitude~\cite{BCJLoop}.

However, for the other gauge-theory copy, ${\cal N} < 4$ sYM, the
three-loop amplitudes are not known. Full-color amplitudes (including
nonplanar terms) are required, and they should be known at the level of
the integrand, because the BCJ form for the three-loop $\NeqFour$ sYM
amplitude contains loop-momentum dependence in its numerator factors.  BCJ
duality for ${\cal N} < 4$ sYM could help simplify these gauge-theory
calculations. For instance, for the three-loop four-point $\NeqFour$ sYM
amplitude, the duality reduced the computation of the full amplitude to
the evaluation of the maximal cut~\cite{FiveLoopYM} of a single
diagram~\cite{BCJLoop}. Non-maximal amplitude calculations are not
expected to be as simple, however.  More powers of loop momentum will
appear in the numerator factors, and graphs containing triangle and bubble
subgraphs will also arise. It would be interesting nonetheless to
investigate the simplifications that may be provided by BCJ duality in
these cases.
 

\section*{Acknowledgments}

We would like to thank Zvi Bern especially, for crucial observations and
extremely helpful suggestions that led to the work performed in this
paper.  We are also grateful to Guillaume Bossard, John Joseph Carrasco and
Henrik Johansson for very stimulating discussions.  
We thank Zvi Bern, John Joseph Carrasco and Henrik Johansson for
insightful comments on the manuscript.  L.D. thanks the CERN theory
group and the Department of Energy's Institute for Nuclear Theory at the 
University of Washington for hospitality while portions of this work
were carried out.  This research was supported by
the US Department of Energy under contract DE--AC02--76SF00515. CBV is
also supported in part by a postgraduate scholarship from the Natural
Sciences and Engineering Research Council of Canada. The figures were
drawn using Jaxodraw~\cite{Jaxo1and2}, based on Axodraw~\cite{Axo}.


\appendix

\section{One-loop expressions}
\label{OneLoopAppendix}

In this appendix we give the $\Ord(\e^0)$ and $\Ord(\e^1)$ coefficients
in the expansion of the one-loop four-graviton
amplitude ${\cal M}_4^{\oneloop}$ in the various supergravity theories,
because they enter the extraction of the two-loop finite remainder
$F_4^{\twoloop}$ according to \eqn{IRexp}.
These amplitudes were first computed through $\Ord(\e^0)$ in
ref.~\cite{DunbarNorridge} for $\NeqFour$ and $\NeqSix$ supergravity
(and the $\NeqFive$ case is trivially related to $\NeqSix$ at one loop).
Expressions valid to all orders in $\e$, in terms of box, triangle and
bubble integrals, can be found in ref.~\cite{Nle8BCJ}.

We write
\be
{\cal M}_4^{\oneloop} = \Bigl(\frac{\ka}{8\pi}\Bigr)^2
\, \left( \frac{4\pi \, e^{-\gamma} \, \mu^2}{|s|} \right)^\e
\, {\cal M}_4^{\tree} \biggl[ 
\frac{2}{\e} \, \Bigl( s \, \ln(-s) + t \, \ln(-t) + u \, \ln(-u) \Bigr)
\, + \, F_4^{(1)} \biggr] \,,
\label{oneloopIRexp}
\ee
where $\ln(-s) \to \ln|s| - i\pi$ in the $s$ channel,
$\ln(-t) \to \ln|t| - i\pi$ in the $t$ channel.
We will give the $\Ord(\e^0)$ and $\Ord(\e^1)$ coefficients for
$F_4^{(1)}$ for each theory in these two channels.

For $\NeqEight$ supergravity in the $s$ channel we have,
\be
F_4^{\oneloop,\,\NeqEight} \Big|_{s-{\rm channel}}
= s \, \Bigl[ g_s\Bigl(\frac{-t}{s}\Bigr)
              + g_s\Bigl(\frac{-u}{s}\Bigr) \Bigr] \,,
\label{NeqEight1ls}
\ee
where
\bea
g_s(x) &=& 2 \, x \, ( \ln x + i\pi ) \, \ln(1-x)
\nonumber\\
&&\null
+ \e \, \biggl\{ - 2 \, (2-x) \, 
               \biggl[ \polylog_3(x) - \frac{\zeta_3}{3}
                + ( \ln x + i\pi ) \, \polylog_2(1-x)
                + \frac{1}{2} \, \ln(1-x) \, ( \ln^2 x - 4 \, \zeta_2 ) \biggr]
\nonumber\\
&&\hskip0.5cm \null
       + \frac{1}{3} \, x \, \ln^3 x
       - i\pi \, (1-x) \, ( \ln^2 x - 4 \, \zeta_2 )
       - \ln x \, ( \ln x + i\pi ) \, \ln(1-x) \biggr\} \,.
\label{gs}
\eea
The $t$-channel result for the $\NeqEight$ supergravity amplitude,
divided by the tree, is obtained by exchanging $s$ and $t$ in the
corresponding $s$-channel result. (This is not quite the case
for $F_4^{\oneloop,\,\NeqEight}$, due to the explicit factor of
$|s|^{-\e}$ extracted in \eqn{oneloopIRexp}.)

We express the finite remainders for ${\cal N} < 8$ supergravities in terms
of the one for $\NeqEight$ supergravity.  For $\NeqSix$ supergravity we find,
in the $s$ channel,
\be
F_4^{\oneloop,\,\NeqSix} \Big|_{s-{\rm channel}}
= F_4^{\oneloop,\,\NeqEight} \Big|_{s-{\rm channel}}
 + s \, \biggl[ g_{6,s}\Bigl(\frac{-t}{s}\Bigr)
              + g_{6,s}\Bigl(\frac{-u}{s}\Bigr) \biggr] \,,
\label{NeqSix1lfinites}
\ee
where
\bea
g_{6,s}(x) &=&
\frac{1}{2} \, x\,(1-x) 
\, \biggl[ \ln^2\left(\frac{x}{1-x}\right) + \pi^2 \biggr]
\label{Neq61ls}\\
&&\null
+ \e \, \biggl\{
  2 \, x\,(1-x) \, \biggl[ \polylog_3(x) - \ln x \, \polylog_2(x)
    - \frac{1}{3} \, \ln^3 x - \frac{\pi^2}{2} \, \ln x \biggr]
\nonumber\\
&&\hskip0.8cm \null
  - \frac{1}{2}
    \, \Bigl[ x \, ( \ln x + i\pi ) + (1-x) \, ( \ln(1-x) + i\pi ) \Bigr]^2
  - \frac{\pi^2}{2} \, ( 1 - x\,(1-x) ) \biggr\} \,.
\nonumber
\eea
The $t$-channel result is
\be
F_4^{\oneloop,\,\NeqSix} \Big|_{t-{\rm channel}}
= F_4^{\oneloop,\,\NeqEight} \Big|_{t-{\rm channel}}
 + s \, g_{6,t}\Bigl(\frac{-u}{t}\Bigr) \,,
\label{NeqSix1lfinitet}
\ee
where
\bea
g_{6,t}(x) &=& 
  - \frac{x}{(1-x)^2} \, \ln x \, ( \ln x + 2 \, i\pi )
+ \e \, \biggl\{
 \frac{2\,x}{(1-x)^2} \, \biggl[ \polylog_3(x) - \zeta_3
               - ( \ln x + i\pi ) \, ( \polylog_2(x) - \zeta_2 )
\nonumber\\
&&\hskip0.8cm\null
               + \frac{1}{3} \, \ln^3 x + 2 \, \zeta_2 \, \ln x
               + \frac{i\pi}{2} \, \ln^2 x
               - \ln x \, ( \ln x + 2 \, i\pi ) \, \ln(1-x) \biggr]
\nonumber\\
&&\hskip0.8cm\null
- \biggl[ ( \ln(1-x) + i\pi ) + \frac{x}{1-x} \, ( \ln x + i\pi ) \biggr]^2
- \frac{1 - x\,(1-x)}{(1-x)^2} \, \pi^2 \biggr\} \,.
\label{Neq61lt}
\eea

The corresponding one-loop results for $\NeqFive$ supergravity
are trivially related to those for $\NeqSix$, because the difference
in field content from $\NeqEight$ is due to the same matter multiplet,
just three copies instead of two.  Therefore we have,
\bea
F_4^{\oneloop,\,\NeqFive} \Big|_{s-{\rm channel}}
&=& F_4^{\oneloop,\,\NeqEight} \Big|_{s-{\rm channel}}
 + \frac{3}{2} \, s \, \biggl[ g_{6,s}\Bigl(\frac{-t}{s}\Bigr)
                             + g_{6,s}\Bigl(\frac{-u}{s}\Bigr) \biggr] \,,
\label{NeqFive1lfinites}\\
F_4^{\oneloop,\,\NeqFive} \Big|_{t-{\rm channel}}
&=& F_4^{\oneloop,\,\NeqEight} \Big|_{t-{\rm channel}}
 + \frac{3}{2} \, s \, g_{6,t}\Bigl(\frac{-u}{t}\Bigr) \,.
\label{NeqFive1lfinitet}
\eea

The $s$-channel one-loop finite remainder for $\NeqFour$ supergravity
is given by
\be
F_4^{\oneloop,\,\NeqFour} \Big|_{s-{\rm channel}}
= F_4^{\oneloop,\,\NeqEight} \Big|_{s-{\rm channel}}
 + s \, \biggl[ g_{4,s}\Bigl(\frac{-t}{s}\Bigr)
              + g_{4,s}\Bigl(\frac{-u}{s}\Bigr) \biggr] \,,
\label{NeqFour1lfinites}
\ee
where
\bea
g_{4,s}(x) &=& \Bigl[ 2 - x\,(1-x) \Bigr] \, g_{6,s}(x)
         + x \,(1-x) \, \biggl[ (1-2\,x) \, \ln x + \frac{1}{2} \biggr]
\nonumber\\
&&\null
- \frac{\e}{6} \, \biggl\{
     \frac{ x \, \bigl( 3 - x^2 \, ( 12 - 15\,x + 5\,x^2 ) \bigr) }{1-x} 
          \, \ln^2 x
   - 5 \, x^2 \, (1-x)^2
         \, \Bigl[ \ln x \, \ln(1-x) - \frac{\pi^2}{2} \Bigr]
\nonumber\\
&&\hskip0.8cm\null
   + i\pi \, \biggl[ 2 \, \frac{x^2}{1-x} \, ( 7 - 12\,x + 6\,x^2 ) \, \ln x
                   + 1 \biggr]
\nonumber\\
&&\hskip0.8cm\null
   - 2 \, x \, ( 6 -  24\,x + 17\,x^2 ) \, \ln x
   - 10 \, x\,(1-x) \biggr\} \,.
\label{NeqFour1ls}
\eea
The $t$-channel expression is
\be
F_4^{\oneloop,\,\NeqFour} \Big|_{t-{\rm channel}}
= F_4^{\oneloop,\,\NeqEight} \Big|_{t-{\rm channel}}
 + s \, g_{4,t}\Bigl(\frac{-u}{t}\Bigr) \,,
\label{NeqFour1lfinitet}
\ee
where
\bea
g_{4,t}(x) &=& \biggl[ 2 + \frac{x}{(1-x)^2} \biggr] \, g_{6,t}(x)
    - \frac{x}{(1-x)^2} \, \biggl[ \frac{1+x}{1-x} \, ( \ln x + i\pi )
                                 + 1 \biggr]
\nonumber\\
&&\null
+ \frac{\e}{6} \, \frac{x}{(1-x)^2} \, \biggl\{
   (3-x\,(1-x)) \, \biggl[ \ln^2\left(\frac{x}{1-x}\right) + \pi^2 \biggr]
    - \frac{5\,x}{(1-x)^2} \, \ln x \, ( \ln x + 2 i \pi )
\nonumber\\
&&\hskip0.8cm\null
    + \frac{1-x+3\,x^2}{x^2} \, \ln(1-x) \, ( \ln(1-x) + 2 i \pi )
    + 2 \, \frac{1-12\,x-6\,x^2}{x\,(1-x)} \, ( \ln x + i\pi )
\nonumber\\
&&\hskip0.8cm\null
    - 2 \, \frac{1-5\,x+x^2}{x} \, \ln\left(\frac{x}{1-x}\right)
    - 20 \biggr\} \,.
\label{NeqFour1lt}
\eea
%


\end{document}